\documentstyle[11pt]{article}

\begin{document}
\large

\def\lsim{\mathrel{\rlap{\lower3pt\hbox{\hskip0pt$\sim$}}
    \raise1pt\hbox{$<$}}}         %less than or approx. symbol
\def\gsim{\mathrel{\rlap{\lower4pt\hbox{\hskip1pt$\sim$}}
    \raise1pt\hbox{$>$}}}         %greater than or approx. symbol
\def\dblint{\mathop{\rlap{\hbox{$\displaystyle\!\int\!\!\!\!\!\int$}}
    \hbox{$\bigcirc$}}}
\def\ut#1{$\underline{\smash{\vphantom{y}\hbox{#1}}}$}

\newcommand{\beq}{\begin{equation}}
\newcommand{\eeq}{\end{equation}}
\newcommand{\dem}{\Delta M_{\mbox{B-M}}}
\newcommand{\dega}{\Delta \Gamma_{\mbox{B-M}}}

\newcommand{\ind}[1]{_{\begin{small}\mbox{#1}\end{small}}}
\newcommand{\WA}{{\em WA}}
\newcommand{\SM}{Standard Model }
\newcommand{\QCD}{{\em QCD }}
\newcommand{\KM}{{\em KM }}
\newcommand{\hscale}{\mu\ind{hadr}}

\newcommand{\ie}{{\em i.e. }}
\newcommand{\eg}{{\em e.g. }}
\newcommand{\etc}{{\em etc. }}
\newcommand{\viz}{{\em viz. }}
\newcommand{\rhs}{{\em rhs }}

\newcommand{\appa}{\mbox{\ae}}
\newcommand{\CP}{{\em CP}}
\newcommand{\fy}{\varphi}
\newcommand{\hi}{\chi}
\newcommand{\al}{\alpha}
\newcommand{\as}{\alpha_s}
\newcommand{\gf}{\gamma_5}
\newcommand{\gm}{\gamma_\mu}
\newcommand{\gn}{\gamma_\nu}
\newcommand{\be}{\beta}
\newcommand{\ga}{\gamma}
\newcommand{\de}{\delta}
\renewcommand{\Im}{\mbox{Im}}
\renewcommand{\Re}{\mbox{Re}}
\newcommand{\GeV}{\,\mbox{GeV}}
\newcommand{\MeV}{\,\mbox{MeV}}
\newcommand{\matel}[3]{\langle #1|#2|#3\rangle}
\newcommand{\state}[1]{|#1\rangle}
\newcommand{\ra}{\rightarrow}
\newcommand{\ve}[1]{\vec{\bf #1}}

%\vspace*{.7cm}
\begin{flushright}
\large{
UND-HEP-93-BIG\hspace*{0.1em}01}\\
TPI-MINN-93/12-T \\
UMN-TH-1149/93\\
March 1993
\end{flushright}
\vspace*{.5cm}
\begin{center} \LARGE {QCD PREDICTIONS FOR LEPTON SPECTRA IN INCLUSIVE
HEAVY FLAVOUR DECAYS}
\end{center}
\vspace*{.3cm}
\begin{center} \Large
I.I.Bigi$^{a,}$\footnote{e-mail address:  BIGI@UNDHEP,
UNDHEP::BIGI},
M.Shifman$^{b,}$\footnote{e-mail address:
SHIFMAN@UMNACVX},
N.G.Uraltsev$^{a,c,}$\footnote{e-mail
address:  URALTSEV@UNDHEP ;  URALTSEV@LNPI.SPB.SU},
A.Vainshtein$^{b,d,}$\footnote{e-mail address:
VAINSHTE@UMNACVX  }
\\

{\normalsize\it $^a$ Dept.of Physics,
University of Notre Dame du
Lac, Notre Dame, IN 46556}
\\
\baselineskip=6pt
{\normalsize\it $^b$ Theoretical Physics Institute,
University of Minnesota, Minneapolis, MN 55455}
\\{\normalsize\it $^c$ St.Petersburg Nuclear Physics Institute,
Gatchina, St.Petersburg 188350, Russia\footnote{permanent address}}
\\{\normalsize\it $^d$ Budker Institute of Nuclear Physics,
Novosibirsk 630090, Russia}
\baselineskip=12pt

\end{center}
\thispagestyle{empty} \vspace{.4cm}

\centerline{\Large\bf Abstract}
\normalsize
\noindent
We derive the lepton spectrum in semileptonic beauty decays from
a nonperturbative treatment of QCD; it is based on an expansion in
$1/m_Q$ with $m_Q$ being the heavy flavour quark mass. The leading
corrections arising on the $1/m_Q$ level are completely expressed in
terms of the difference in the mass of the heavy hadron and the quark.
Nontrivial effects appear in $1/m_Q^2$ terms affecting mainly the endpoint
region; they are different for meson and baryon decays as well as
for beauty and charm decays.

\thispagestyle{empty}

\vspace*{.85cm}

\large
\addtocounter{footnote}{-5}

The weak decays of hadrons contain
a wealth of information on the fundamental forces of nature.
Yet the intervention of the strong interactions has prevented us from
extracting this
information in a reliable way. Heavy flavour decays promise
to be more tractable since the mass of the heavy flavour
quark $m_Q$ provides a powerful expansion parameter.
Indeed significant successes
have been scored by Effective Heavy Quark Theory (EHQT)\cite{EHQT}.
Yet that approach has some intrinsic limitations:
e.g., it deals with exclusive
semileptonic modes only and it requires the presence of
heavy quarks both in the initial and in the final state;
thus it cannot be applied directly to
$b\rightarrow u$ transitions. Its model-independent
predictions so far include corrections through order $1/m_Q$ only.
On the other hand the energy released in heavy
flavour decays is much larger than ordinary hadronic energies.
Our analysis will make use of this large energy release in treating
$Q\rightarrow q l \nu$ transitions with
$m_Q,m_Q-m_q\gg \Lambda _{QCD}$.

In previous papers \cite{we} we have shown how nonperturbative
contributions to
global quantities like lifetimes and semileptonic branching ratios
can be obtained from a model-independent treatment of QCD.
The method was based on expanding the
weak transition operator into a series of local operators of increasing
dimension with coefficients that contain increasing powers of $1/m_Q$.
The coefficients depend on the (inclusive) final state;
the differences between the decay rates of different heavy flavour hadrons
$H_Q$ -- charged vs. neutral mesons vs. baryons -- enter through the
matrix elements of the local operators taken between $H_Q$.
For example the total semileptonic $b \ra u$ width through order
$1/m_b^2$ is given by:
$$\Gamma (H_b\rightarrow l \nu X) \propto m_b^5 \;\cdot
\;\frac{1}{2M_{H_b}}\;
\matel {H_b}{\;\bar b b \;- \;\frac {1}{m_b^2}\bar b\,
i\sigma G \,b}{H_b} \;\; \eqno(1)$$
with $i\sigma G=i\gamma_{\mu}\gamma_\nu G_{\mu \nu}$ ,
$G_{\mu \nu}$ being the gluonic field strength tensor.

In this note we will expand the general method to treat the
lepton {\em spectra} in
the semileptonic decays of heavy flavour hadrons.
A novel feature is encountered:
one is dealing with an expansion in powers of
$1/(p_Q-p_l)^2$ with $p_Q$ and $p_l$ denoting the
momenta of $Q$ and the lepton $l$, respectively. This series is
singular at the endpoint of the lepton spectrum; thus some care
has to be applied in interpreting the results.
Two dimension five operators generate the leading nonperturbative
corrections of order  $1/m_Q^2$ : the colour magnetic operator
$\bar Q\, i\sigma G \,Q$ and the operator
$\bar Q\,(D^2-(v D)^2)\,Q$ describing
the kinetic energy of $Q$ in the gluon background field;
$D_\mu$ denotes here the covariant derivative and $v_{\mu}$ is the
4-velocity vector of the hadron. Corrections actually arise
already on the $1/m_Q$ level; it is crucial that in QCD those
 can be expressed completely in terms of the difference
between the quark and the hadron mass.

We will phrase our discussion in terms of beauty decays
with a few added comments on charm decays.
It should be noted that the question of the inclusive
lepton spectra in QCD was first addressed explicitely in ref. \cite{TM}.

Ignoring gluon bremsstrahlung one obtains a lenghty expression
for the lepton spectra in the semileptonic decays
of a {\em beauty hadron} $H_b$:
$$ \frac {d\Gamma }{dy}(H_b\rightarrow l\nu X_q) \;= \Gamma_0\;
 \theta (1-y-\rho)\,2y^2\{(1-f)^2
(1+2f)(2-y)+(1-f)^3(1-y) + $$
$$+(1-f)\;
[\;(1-f)(2+\frac {5}{3} y -2 f + \frac{10}{3} f y )\;-\;
\frac {f^2}{\rho}(2y+f(12-12y+5y^2))\;] \,G_b\;-$$
$$\;-\; [\;\frac {5}{3}(1-f)^2(1+2f) y \,+\,
\frac {f^3}{\rho}(1-f)(10y-8y^2)\,+\,
\frac {f^4}{\rho ^2}(3-4f)(2y^2-y^3)\:] \,K_b\;\}$$
$$ \Gamma_0=\frac{G_F^2m_b^5}{192\pi^3}|V_{qb}|^2\;\;,\;\;\;\;
f=\frac {\rho}{1-y}\;,\; \;\;\;
\rho = \frac {m_q^2}{m_b^2}\;,\; \;\;\; y=\frac {2E_l}{m_b}\;,$$
$$K_b=1/2M_{H_b}\cdot\matel {H_b}{\bar b (i\vec{D})^2b}{H_b}/m_b^2\;\;\; ,
\;\;\
;\;\;\;\;
G_b=1/2M_{H_b}\cdot\matel {H_b}{\bar b\,i\sigma G\,b}{H_b}/2m_b^2 \eqno(2)$$
with $m_q$ denoting the mass of the quark $q$ in the final state.
Eq.(2) represents the master formula containing both relevant
cases, namely $q=u,\, c$.

For $b\rightarrow ul \nu$ transitions with $m_u=0$ this
expression simplifies considerably:
$$\frac {d\Gamma }{dy}=\Gamma_0 2y^2[3-2y
-(\frac{5}{3}y+\frac {1}{3}\delta (1-y)-\frac{1}{6}(2y^2-y^3)
\,\delta '(1-y))K_b+
(\,2+\frac{5}{3}y-\frac {11}{6}\delta (1-y))\:G_b\:]\;\;\eqno(3)$$
The $\delta$-functions and their derivative
reflect the previously mentioned singular nature of the expansion at
the endpoint; their emergence has a transparent meaning (see\cite{next} for
details).
The spectrum is finite at the endpoint for $m_u=0$ and thus contains a
step function.
The chromomagnetic interaction effectively `shifts' the
spectrum by changing the energy in either initial or final state;
the shift in the argument of the step function
thus yields a $\delta$-function. The singular structure in the $K_b$
term on the
other hand reflects the motion of the b quark inside
the $H_b$ hadron which Doppler shifts the spectrum; in second
order it generates $\delta^\prime (1-y)$ for the step-like spectrum.

Due to these singular terms the expression given above can be
identified with the observable spectrum only {\em outside a finite
neighbourhood of the endpoint region}. (This distance remains constant in
absolute units in the limit $m_b\ra \infty$). Yet even this neighbourhood
does not represent true `terra incognita': for integrating our
expression over this kinematical region yields a finite and trustworthy
result that can be confronted with the data.
This can be expressed through the function
$$ \Gamma (E_l)=\int ^{E_{max}}_{E_l}dE_l\;\frac{d\Gamma} {dE_l}\;\;,
\;\; \; \; E_l \le E_{QCD}<E_{max}\;\;\;.\eqno(4)$$
$E_{max}$ denotes the maximal kinematically allowed energy and
$E_{QCD}$ the maximal energy for which one can still trust the QCD
expansion in eqs.(2,3); its value depends on the size of $K_b$ and
$G_b$.
Clearly  $\Gamma (0)=\Gamma _{SL}$ has to hold. This is not a trivial
relation: for $\Gamma _{SL}$ is deduced from a
completely regular expansion in $1/m_b\,$ (see eq.(1))
whereas $\Gamma (0)$ comes from integrating the
expression in eq.(3) containing singularities; thus the singular terms are
essential for recovering the correct decay width!

Similar considerations apply to $b\ra cl\nu$
transitions. With $m_c$ as an infrared
cutoff there arise no singular terms at the endpoint
$y=1-\rho$; yet the expansion parameters $G_b/(1-y), K_b/(1-y)^2$
though finite become large there.
The need for `smearing' the spectrum over the endpoint
region, eq.(4), thus still exists.

The contributions from higher dimensional operators that we are
ignoring have terms of the
schematical form
$\sim [\hscale /(m_b(1-y))]^n$; summing them all up
would yield a well-behaved function.
As long as these quantities are smaller
than unity, we can trust the expressions given above for the
unintegrated lepton spectrum.

The size of $G_b$ is easily determined:
$G_b= \matel {B}{\bar Q\,
i\sigma G Q\,}{B} /4m_b^3 \simeq 3(M^2(B^*)-M^2(B))/4m_b^2\simeq 0.017$;
for $\Lambda_b$ it vanishes.
A recent QCD sum rules analysis yields \cite{VLB}
$ \matel {B}{\bar b\, (i\vec {D})^2\,b}{B}
\sim 0.4\GeV^2\cdot 2M_B$ in agreement
with rather general expectations. For our subsequent discussion we
will set
$ K_b = 0.02$.
We use $m_b\simeq 4.8 \GeV$
as deduced from a QCD analysis of the Ypsilon system,
$m_b-m_c=3.35 \GeV$ as inferred
from the beauty and charm meson masses which for $m_b=4.8\GeV$ yields
$m_c=1.45 \GeV$, and put $m_u=0$.
For $b\rightarrow u$ decays we estimate that $E_{QCD}\sim 0.9 \cdot m_b/2
\simeq 2.15\GeV\,$; for $b\rightarrow c$ a
somewhat smaller `smearing' range near the endpoint seems to be
required, namely $\Delta E_l\simeq 0.15\GeV$ and thus
$E_{QCD}\simeq 2.0\GeV$. Numerically it implies that in the real world the
$c$ quark is
relatively heavy in $b$ decays: $m_c^2 > \hscale m_b\,$; therefore the
falling edge of the spectrum starts in the calculable region.

For a proper perspective
we show the partially integrated spectrum $\Gamma (E_l)$
from three different prescriptions,
namely {\bf (a)} our QCD expansion; {\bf (b)} the simple free
quark picture with $G_b=K_b=0,\, m_u=0$; {\bf (c)} the
phenomenological treatment of Altarelli et al. \cite{ACM}
(hereafter referred to as ACM)
where one attempts to incorporate some bound state effects. We have
set here $m_u=m_{spect}=0.15 \GeV$, $m_c=1.67\GeV$,
 $p_F=0.3\GeV$  as suggested by a fit to CLEO data;
$p_F$ denotes the `Fermi momentum'.
To be consistent we have ignored gluon bremsstrahlung both in ACM and in
our QCD expressions (for fitting the data it should be added
to eqs.(2-4)). In comparing the QCD formula with the ACM
prediction one has to note a subtle distinction
in the definition of the
kinematical variables:
in ACM energy is expressed in terms of the
mass of the beauty {\em hadron}; yet in eqs.(2-3) $y$ measures the energy in
units of $m_b$, which by itself introduces a shift of order $1/m_b$.

Comparing the results on $\Gamma (E_l)$ from the three approaches
we conclude:
{\bf (i)} The {\em shape} of the QCD, of the free quark model
and of the ACM
curves are very close over most of the range of
$E_l \le E_{QCD}$ for the $b\rightarrow u$ as well as the
$b\rightarrow c$ case.
The main difference lies in the overall {\em normalization}.
{\bf (ii)} The QCD result for $b\rightarrow u$ can be largely simulated
by setting $m_u\sim 0.3\, GeV$ in
the free quark spectrum. The nonperturbative
corrections thus effectively transform a
current quark mass into a `constituent' one of reasonable size.
{\bf (iii)} Once the ACM result is renormalized according to
$\;\Gamma (0)\ind{QCD} (b\rightarrow u)=
1.07\times \Gamma (0)\ind{ACM}(b\rightarrow u)\;$ and
$\;\Gamma (0)\ind{QCD}(b\rightarrow c)=
1.13\times \Gamma (0)\ind{ACM}(b\rightarrow c)\;$
the two curves are hardly distinguishable as functions of $E_l$.
{\bf (iv)} While the QCD results for
$\Gamma (E_l)$ are largely independent of
the value of $K_b$ for $K_b\sim 0.02$ in
$b\rightarrow u$, they are sensitive to it for $b\rightarrow c$; that change
can be easily understood\cite{next}.
{\bf (v)} The QCD curves for semileptonic $\Lambda _b$ decays are somewhat
harder than for $B$ meson decays.
{\bf (vi)} The largest differences between the models are found in
the endpoint region, namely for
$E_l>1.8 \GeV$ in $b\rightarrow c$.

Some of these points are illustrated in the figures. In fig.1
we show $\Gamma (E_l)$ for the endpoint region of $b\rightarrow u$;
in fig.2 we have plotted these partially integrated
spectra for $b\rightarrow c$.
The {\em unintegrated} spectrum $d\Gamma(E_L) /dE_l$
which can be calculated in the QCD treatment for
$E_l \leq 2\GeV$ is shown in fig.3.
The similarity in the shape of the two spectra is quite striking!

A few comments are in order about charm decays.
The lepton spectra
are quite different
in beauty and in charm decays, especially in the endpoint region:
for the charged lepton in beauty decays is an electron or muon
whereas it is an antifermion $l^+$ in charm decays.
Since the lepton spectrum vanishes at the end point in
$c\rightarrow ql\nu$ even for $m_q=0$,
the chromomagnetic operator can yield finite terms only while
the kinetic energy operator produces a $\delta (1-y)$, but
not a $\delta ^{\prime}(1-y)$ term.
Yet since the nonperturbative corrections are much larger in charm
than in beauty -- $\;G_c,K_c \sim 0.2\;$ --
the smearing required in $y$ is much larger for charm, namely
$\Delta y \sim 0.5$. While this allows us meaningful statements
about fully integrated quantities like $\Gamma\ind{SL}$, it appears
beyond the scope of our present analysis to treat lepton
{\em spectra} in charm decays.
\vspace*{.1cm}

\noindent
{\Large \em To summarize: \hspace*{1em}}
We have shown here how an
expansion in $1/(p_b-p_l)$ allows us to incorporate
nonperturbative corrections to the lepton spectra in
inclusive semileptonic $b\rightarrow c$ and $b\rightarrow u$
decays. Due to the singular
nature of this expansion at the endpoint the spectrum cannot be
computed in detail in a close neighbourhood of the endpoint, yet smeared
or partially integrated
spectra can. Through order $1/m_b^2$ these expressions are given
in terms of the quantities $m_b$, $m_c$, $G_b$ and $K_b$. These are
{\em not} free parameters -- their size can be extracted from the
relationship to other observables. Improvements in the precision of
their determination, in particular concerning $K_b$ and to a lesser
degree $m_b$, can be anticipated from future progress in theory.

We have compared our QCD results with the ACM description that
so far has allowed a decent fit to the data. Despite the obvious
differences in the underlying dynamics
we have found the shape of the resulting lepton spectra
remarkably similar, and even more so when one keeps the following in
mind:

\noindent {\bf (a)} Initially the scale for the kinematics is
different in the two descriptions:
for the QCD expansion it is set by the quark mass $m_b$ whereas for ACM by
the
hadron mass $M_B$ or $M_{\Lambda_b}$. The kinematical differences due to
$m_b\ne M_{H_b}$
are actually of order $1/m_b$ and formally represent the
leading corrections.  This suggests an interesting
observation: an accurate measurement of the shape of the spectrum
allows a determination of the $b$ quark mass $m_b$ free of theoretical
uncertainties.
Unfortunately in practice one would have to analyse $b \ra u$; the shape of
the $b \ra c $ spectrum is basically determined by $m_b-m_c$ with little
sensitivity to $m_b$.

\noindent {\bf (b)} The ACM ansatz contains three free parameters
-- $m_c,\, m_{sp}$ and $p_F$ -- that are to be fitted from the data. There
are four quantities that set the scale in the QCD expansion through order
$1/m_b^2$, namely $m_b,\, m_c,\, G_b$ and $K_b$ but none of them is a free
parameter. That our QCD description
containing therefore very little `wiggle room' can nevertheless yield a
description so similar to that of the ACM ansatz -- after the latter has been
fitted to the data -- has to be seen as quite remarkable!

\noindent {\bf (c)} The ACM prescription can thus be viewed
as a simple, though smart approximation to a more complete and complex QCD
treatment.
At the same time it would be incorrect to interprete the fit parameters in
ACM literally as real physical quantities; it is thus not surprising that the
numerical values for the former differ significantly from what is known now
about the corresponding quantities in QCD.

A more detailed analysis shows\cite{next} that QCD
provides a natural `home' for the Fermi motion originally introduced
phenomenologically in ACM, and it is asymptotically a dominant effect for the
end point shape in $b$ decays. However it enters in a somewhat different
form. Its impact on the total widths is only
quadratic in $1/m_b$ (in ACM it produces a linear shift).
Nevertheless the {\em shape} of the lepton spectrum gets
$1/m_b$ corrections\footnote{At this point we disagree with
the conclusion of ref.\cite{TM}}; it will be discussed in detail
elsewhere\cite{next}.

Despite all similarities there arise relevant numerical differences as well,
not only in the overall normalization of the spectra, but
also in the endpoint region: the QCD spectra place a {\em higher}
fraction of $b\rightarrow c$
events in the endpoint region. Such a difference
has important consequences for the extraction of
$|V_{ub}/V_{cb}|$ from the data. Furthermore our analysis shows
that the lepton spectra in semileptonic $\Lambda _b$ decays are
distinct from those in  $B$ decays. These issues will
be discussed in a future paper\cite{next}.

Another model ISGW\cite{IW} used for describing
semileptonic spectra relies heavily on model-dependent calculations of the
formfactors for the exclusive final states. In the limit of a
heavy $c$ quark, $m_b-m_c \ll m_c\,$, EHFT
yields the necessary formfactors to some accuracy; therefore in that limit
the prediction of this model coincides with the model-independent QCD
spectrum. On the other hand that is apparently not the case for
$b\ra u$ decays
where the difference is significant.

On the $1/m_Q^3$ level that was not treated here novel effects
arise due to `Weak Annihilation' in the $b\ra u$ channel; for
details see \cite {BU3}.
\newpage
%\vglue 0.3cm
{\Large \bf \noindent Acknowledgements \hfil}
\vglue 0.2cm
This work was supported in part by the NSF under grant
number NSF-PHY 92-13313 and in part by DOE under grant number
DOE-AC02-83ER40105.

\vglue 0.3cm
{\Large \bf\noindent Figure Captions \hfil}
\vglue 0.2cm

{\bf Fig.1:} The function $\Gamma (E_l)$ for $b\rightarrow u$ transitions
calcul
ated
in the QCD expansion (solid line), the free spectator quark model
(dotted line) and the ACM ansatz (dashed line); the ACM curve has been
multiplied by a factor 1.07.

{\bf Fig.2:} The function $\Gamma (E_l)$ for $b\rightarrow c$ transitions
calcul
ated
in the QCD expansion (solid line), the free spectator quark model
(dotted line) and the ACM ansatz (dashed line); the ACM curve has been
multiplied by a factor 1.13.

{\bf Fig.3:} The lepton spectrum $d\Gamma /dE_l$ for
$b\rightarrow c$ transitions calculated
in the QCD expansion (solid line), the free spectator quark model
(dotted line) and the ACM ansatz (dashed line); the ACM curve has been
multiplied by a factor 1.13.

\end{document}